\documentclass[french]{article-hermes}
\usepackage{graphicx}

\bibstyle{biblio-hermes}

\begin{document}
%\maketitle % Exemple de 1ere page sans les valeurs de champs
            % pour afficher les commandes à utiliser
%\journal[Mode d'emploi]{Signature de l'article~: nom de la revue}%
%   {10}{5}{1999}{1}{15}% Exemple avec numéros de page
%\journal{DECOR}{}{}{2004}{1}{2}

\title[Déploiement d'ordonnanceurs]{Déploiement d'ordonnanceurs de processus spécifiques dans un système d'exploitation\\généraliste}

\author{Hervé Duchesne \andauthor Christophe Augier \andauthor Richard Urunuela }

\address{
\'Equipe OBASCO\\
\'Ecole des Mines de Nantes \\
Département informatique\\
4 rue Alfred Kastler\\
F- 44307 Nantes cedex 3\\[3pt]
\{hduchesne,caugier,rurunuel\}@emn.fr}

\resume{L'environnement Bossa permet la conception d'ordonnanceurs spécifiques offrant ainsi une gestion précise de la ressource processeur au sein d'un système d'exploitation généraliste. Toutefois, le déploiement conjoint d'une application et de son ordonnanceur suscite des problèmes liés à la réservation de la ressource processeur. Dans cet article, nous étudions les problèmes dus à ce type de déploiement et proposons quelques solutions. En particulier, nous proposons d'établir des contrats de qualité de service et des mécanismes de reconfiguration de la hiérarchie d'ordonnanceurs.
}
\abstract{Bossa is a framework to develop new processes schedulers in commodity
operating systems. Although Bossa enables fine-grained management of the
processor through new scheduling policies, deploying an
application with its own scheduler raises some problems. In this paper we
study the problems caused when deploying an application and its
scheduler and to adresse these, we propose to establish Quality of Service contracts and mechanisms to reconfigure the scheduler hierarchy.
}

\motscles{déploiement d'ordonnanceurs, gestion de ressources}


\proceedings{DECOR'04, Déploiement et (Re)Configuration de Logiciels}{193}

\maketitlepage
%--------------------------------------------%
%INTRO
\section{Introduction}

%*************************** VERSION CORRECTE ********************************%
%Les récents progrès survenus dans le secteur du matériel informatique permettent aux machines actuelles d'exécuter simultanément plusieurs applications de classes hétérogènes telles que des applications serveurs et des applications multimédia.
%Les applications multimédia comme celles permettant le visionnage de vidéos ou la décompressions synchrone de fichiers audios sont de plus en plus prisées par des utilisateurs qui les exécutent sur des systèmes d'exploitations généralistes. Or les applications multimédia doivent répondre à des contraintes de qualités de service.
Les applications multimédia sont de plus en plus prisées par les utilisateurs. Ces applications sont le plus souvent exécutées sur des systèmes d'exploitations généralistes et ont la particularité de devoir répondre à de fortes contraintes de qualité de service. Le respect de ces contraintes dans un système d'exploitation nécessite une gestion efficace de la ressource processeur par l'ordonnancement de processus. Cependant, dans les systèmes d'exploitation généralistes, les ordonnanceurs natifs ne sont pas en mesure de garantir, par exemple, le respect des contraintes temporelles nécéssaires à la bonne exécution de certaines applications multimédia~\cite{KG00,HDR}. %\cite[].
%Afin de garantir le respect de ces contraintes temporelles
Une solution à ce problème est de développer des ordonnanceurs spécifiques pour cette classe d'applications. Toutefois, le développement d'un nouvel ordonnanceur requiert une bonne connaissance des noyaux de systèmes d'exploitation et des langages de programmation bas niveau tels que le C ou l'assembleur. Peu de programmeurs possèdent ces deux expertises.
%Or acquérir ce type de compétence n'est pas facile.

%Une approche par langage dédié (DSL) permet de dépasser ces difficultés an facilitant la conception d'ordonnanceurs spécifiques à une application ou un ensemble d'applications partageant les même besoins.

L'environnement Bossa a été développé pour faciliter la conception et l'intégration de nouveaux ordonnanceurs au sein d'un système d'exploitation généraliste~\cite{RTS02,GPCE04}. Bossa permet d'associer un ordonnanceur spécifique à chaque application ou ensemble d'applications partageant les mêmes besoins~\cite{PEPM}. Cependant, le déploiement conjoint d'une application et de son ordonnanceur suscite des problèmes liés à l'expression des besoins d'ordonnancement de l'application et du multiplexage de la ressource processeur entre les différents ordonnanceurs chargés. Pour adresser ces problèmes, nous proposons de mettre en place un protocole de déploiement d'ordonnanceurs. \`A cette fin,  nous étudions sous quelles conditions l'intégration d'un ordonnanceur spécifique est requise et quelle est l'influence du déploiement d'un ordonnanceur sur le système.

%Cependant, la coexistance de plusieurs ordonnanceurs au sein d'un même système pose le problème du partage de la ressource processeur entre ces différents composants d'ordonnancement. En effet, la ressource processeur n'étant pas extensible,

%l'intéret de
%En effet, la ressource processeur n'étant pas extensible, il faut dans un premier temps valider la pertinance du chargement d'un ordonnanceur et dans un second temps, étudier l'influence de l'intégration d'un ordonnanceur au sein du système. Cette problématique a été assez peu étudiée.
% ce déploiement conjoint d'une application et de son ordonnanceur du fait que nous disposions pas d'environnement comme Bossa permettant  .

La section 2 est une brève introduction aux concepts de l'environnement Bossa. La section 3 présente les problèmes engendrés par le déploiement d'une application avec son ordonnanceur. Quelques solutions à ces problèmes sont présentées en section 4. La section 5 conclut cet article et présente quelques perspectives.

\section{L'environnement Bossa}

%****************************** VERSION CORRECTE *******************************%
L'environnement Bossa se situe dans le domaine des extensions de systèmes d'exploitation où l'on trouve des travaux comme SPIN~\cite{SPIN} ou Flux OSKit~\cite{FLUX}. Aussi, Bossa permet de développer et d'intégrer facilement et de façon sûre des ordonnanceurs de processus dans un système d'exploitation.

\begin{figure}[tb]
%\begin{center}
\includegraphics[angle=0, scale=0.35]{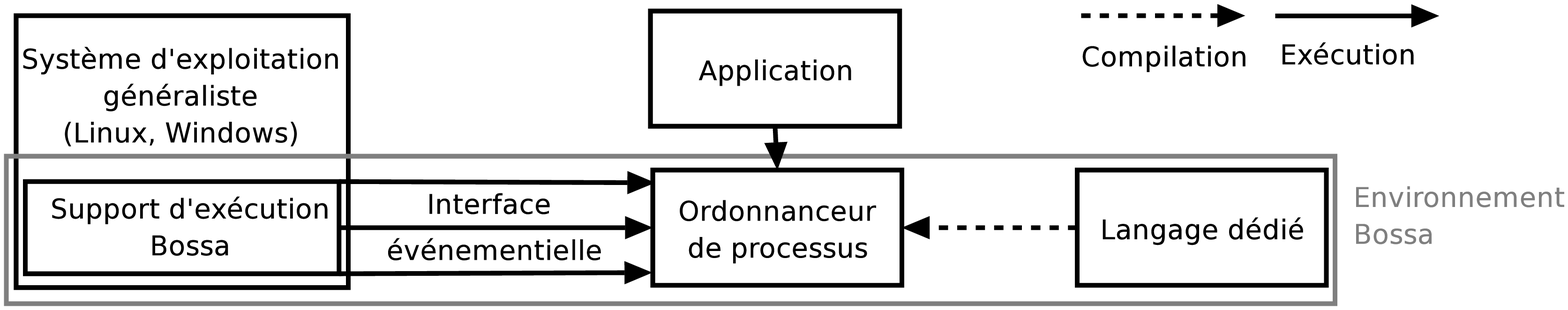}
\caption{Environnement Bossa}
\label{fig:EnvBossa}
%\end{center}
\end{figure}

L'environnement Bossa est organisé autour d'un langage dédié (\textit{domain specific language} DSL) et d'un support d'exécution, comme le montre la Figure~\ref{fig:EnvBossa}. Le DSL offre au programmeur des abstractions de haut niveau dédiées au domaine de l'ordonnancement et facilite ainsi l'implémentation de nouvelles politiques d'ordonnancement. Le support d'exécution fournit au système d'exploitation une interface facilitant l'intégration de composants d'ordonnancement. Dès lors, chaque classe d'applications exécutées sur une même machine peut disposer de son ordonnanceur spécifique.
%L'environnement Bossa permet d'associer un ordonnanceur spécifique à une application ou un ensemble d'applications partageants les même besoins. Ainsi,
% résumé sous le terme de classe d'applications.
\label{bossa}

\begin{figure}[htb]
%\begin{center}
\includegraphics[angle=0, scale=0.35]{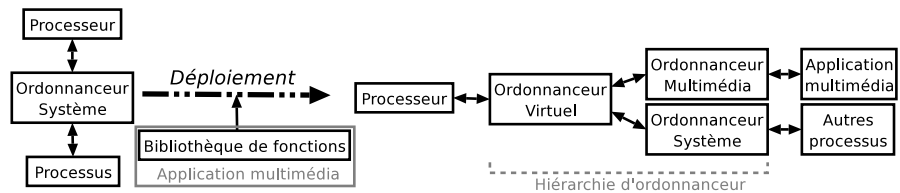}
\caption{Exemple du déploiement actuelle d'une application et son ordonnanceur dans l'environnement Bossa}
\label{fig:HieBossa}
%\end{center}
\end{figure}

Toutefois, l'exécution simultanée d'ordonnanceurs au sein d'un même système soulève des problèmes de multiplexage de la ressource processeur. Aussi, pour gérer l'accès des différents ordonnanceurs à la ressource processeur, il peut s'avérer nécessaire de déployer une entité de régulation. Dans l'environnement Bossa, cette entité de régulation est implémentée sous la forme d'un ordonnanceur virtuel qui peut être vu comme un ordonnanceur d'ordonnanceurs~\cite{PEPM}. L'association d'un ordonnanceur virtuel et d'un ou plusieurs ordonnanceurs constitue une hiérarchie d'ordonnanceurs comme le montre la Figure \ref{fig:HieBossa}. La construction d'une hiérarchie, dans l'environnement Bossa, se fait de façon explicite. Aussi, une application est en mesure de déployer son propre ordonnanceur via l'utilisation d'une bibliothèque de fonctions ad-hoc, comme le montre la Figure \ref{fig:HieBossa}.
%Le déploiement d'une application avec son ordonnanceur est possible dans l'environennement Bossa. En effet, une bibliothèque de fonctions ad-hoc permet à une application de charger son ordonnanceur comme le montre la figure \ref{fig:HieBossa}. La fonction \verb?mount_root()? charge l'ordonnanceur virtuel servant à multiplexer la ressource processeur avec l'ordonnanceur existant et la fonction \verb?mount()? charge l'ordonnanceur de l'application.

% peut être montée directement par le lecteur vidéo via une bibliothèque de fonctions distribuée avec l'environnement bossa. Ainsi pour déployer un ordonnanceur une application doit contenir deux fonctions, \verb?mount()? et \verb?mount_root()? qui permettent respectivement le chargement de l'ordonnanceur et celui de l'ordonnanceur virtuel.
%qui permet de multiplexer la ressource processeur comme le montre la figure \ref{fig:HieBossa}. Le déploiement de l'ordonnanceur Earliest Deadline First [] est fait par l'application elle même via l'utilisation d'une bibliothèque de fonctions dont \verb?mount()? et \verb?mount_root()? sont les plus utilisées...

\section{Problèmes engendrés par le déploiement d'une application et de son ordonnanceur}

%******************************** VERSION CORRECTE **************************%
%Le déploiement d'une application avec son ordonnanceur se déroule généralement en deux étapes. Dans un premier temps, l'application spécifie l'ordonnanceur qu'elle souhaite utiliser. Dans un second temps, l'ordonnanceur est chargé par le système d'exploitation et l'application lui est attachée.
Le modèle de déploiement d'une application et de son ordonnanceur dans l'environnement Bossa %tel que présenté en section \ref{bossa}
 exhibe les limitations suivantes :
\begin{itemize}
 % Statique
 \item les conditions d'exécution de l'application sont
 % passage de paramètres est
 définies à priori par le programmeur sans tenir compte de l'environnement d'exécution,
 % De plus le déploiement est local car:

\item le système ne vérifie pas la pertinence du chargement d'un nouvel ordonnanceur, c'est-à-dire s'il n'existe pas déjà un ordonnanceur sur lequel l'application est en mesure de s'exécuter,
\item le système ne vérifie pas lors du chargement d'un ordonnanceur que les garanties offertes initialement aux applications existantes sont encore respectées. En effet, l'exécution simultanée de plusieurs ordonnanceurs sur la même ressource processeur entraîne irrémédiablement des problèmes de partage.
\end{itemize}
Pour garantir le déploiement cohérent d'une application et son ordonnanceur avec l'environnement existant, le système doit offrir des mécanismes pour répondre aux limitations actuelles.

\vspace*{-0,25cm}
\section{Solutions aux problèmes de déploiement}
%************************* CHANGER DE PLACE *****************************%
%Pour répondre aux problèmes du déploiement d'une application avec son ordonnanceur, le système doit, lors de chaque exécution de l'application, négocier l'intégration de l'ordonnanceur spécifique dans la hiérarchie et le cas échéant, valider l'adéquation entre les besoins de l'application et la disponibilité de la ressource processeur.
%************************************************************************%

%****************************** NOUVELLE INTRO **************************%
Nous introduisons les notions de classe d'applications et de services d'ordonnancements. Aussi, pour valider la pertinence du chargement d'un nouvel ordonnanceur, nous proposons de mettre en place un protocole de communication %entre une application et la hiérarchie d'ordonnanceurs
~\cite{HDR}. %Ce protocole permet à une application de
Dans ce protocole, une application peut spécifier la classe à laquelle elle appartient, ses besoins en matière d'ordonnancement ainsi que  %offrant ainsi la possibilité
 vérifier l'existance d'un service d'ordonnancement compatible dans la hiérarchie d'ordonnanceurs.
 Pour préserver les garanties d'ordonnancement lors du chargement d'un nouvel ordonnanceur dans une hiérarchie, nous proposons d'utiliser un algorithme de construction de hiérarchies. Pour expliciter ces solutions, nous décrivons tout d'abord une approche introduite par Regehr et Stankovic et proposons ensuite quelques extensions à ce modèle~\cite{Regehr01}.

\vspace*{-0,35cm}
%*************************************************************************%

%Cette partie présente une première approche réalisée par Regehr et Stankovic Dans l'article \cite{Regehr01} pour répondre aux problèmes de déploiement non encore résolus. Par la suite, nous montrons les limites de cette approche et proposons quelques extensions afin de mieux répondre à notre problématique.
%comment et selon quel protocole une application spécifie t'elle l'ordonnanceur qu'elle souhaite utiliser et ii) comment garantir la cohérense globale d'une hiérarchie d'ordonnancement lorsque l'on y intégre un nouvel ordonnanceur.
%Toutefois, il faut, lors de ce déploiement, valider la pertinence du chargement d'un nouvel ordonnanceur et
%il faut vérifier, lors du chargement d'un nouvel ordonnanceur, que len exprimant et communiquant ses besoins d'ordonnancement si l'ordonnanceur demandé n'est pas présent il

%La négociation pour l'intégration d'un ordonnanceur consiste à vérifier, dans la hiérarchie d'ordonnanceur, l'existance  d'un service d'ordonnancement compatible avec les besoins de l'applications.
%Si aucun service n'est trouvé, le système charge l'ordonnanceur spécifique fourni avec l'application.

%\subsection{Caractérisation des services d'ordonnancement}
\subsection{Approche de Regehr et Stankovic}
%***************************** VERSION ENVOYEE A JULIA *******************%
%Pour vérifier la compatiblité entre un service d'ordonnancement et une application, le système doit discriminer les ordonnanceurs en fonction des caractéristiques qu'ils présentent. Dans l'article \cite{Regehr01}, Regehr et Stankovic exposent leur travaux sur la caractérisation des services d'ordonnancement. Ainsi, les auteurs proposent qu'un ordonnanceur établisse un contrat de qualité de service avec l'entité qu'il ordonnance (Thread ou ordonnanceur) selon la syntaxe \verb?TYPE[param]? où \verb?TYPE? caractérise le service d'ordonnancement (ordonnanceur temps réel, à proportion...) et \verb?[param]? représente le quota de ressource processeur garanti par le contrat. L'expression de ce quota et donc le nombre de paramètre, dépend du type du service d'ordonnancement. Ainsi un contrat garantissant localement 40\% de la ressource processeur, passé entre une entité à ordonnancer et un ordonnanceur à proportion s'exprime \verb?PS[40]?.
%*******************************************************************************%

%****************************** NOUVELLE VERSION *******************************%
%Pour vérifier la compatiblité entre un service d'ordonnancement et une application, le système doit discriminer les ordonnanceurs en fonction des caractéristiques qu'ils présentent. Pour cela, il est nécessaire de définir une syntaxe permettant l'expression des besoins d'ordonnancement et  de classifier les ordonnanceurs en fonctions des services qu'ils proposent.
Regehr et Stankovic ont mis en place une caractérisation des services d'ordonnancement qui  permet d'évaluer et de valider l'utilisation de la ressource processeur au sein d'une hiérarche d'ordonnanceurs. %~\cite{Regehr01}.
 Aussi, Regehr et Stankovic proposent qu'un ordonnanceur établisse un contrat de qualité de service avec l'entité qu'il ordonnance (application ou ordonnanceur). Un contrat de service s'exprime selon la syntaxe \verb?TYPE[param]?, où \verb?TYPE? représente le service d'ordonnancement (ordonnanceurs temps réel, à proportion...) et \verb?[param]? représente le quota de ressource processeur garanti par le contrat. La combinaison des différents contrats ébauche le squelette d'une hiérarchie, qui est ensuite construite par un algorithme spécifique.

Ce modèle de déploiement permet de résoudre partiellement i) le problème de négociation de l'intégration d'un ordonnanceur en offrant une classification des ordonnanceurs par type de service d'ordonnancement et ii) le problème de reconfiguration dynamique de la hiérarchie en proposant un algorithme de composition de hiérarchie d'ordonnanceurs.

\vspace*{-0,4cm}
\subsection{Extension des travaux de Regehr et Stankovic}

\begin{figure}[h!]
%\begin{center}
\includegraphics[scale=0.35]{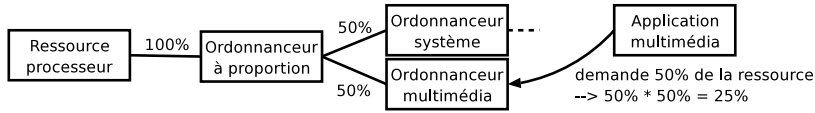}
\caption{Limitation de la ressource processeur au niveau local}
\label{fig:local}
%\end{center}
\end{figure}

\begin{center}
\begin{figure}[h!]
\begin{center}
\includegraphics[scale=0.30]{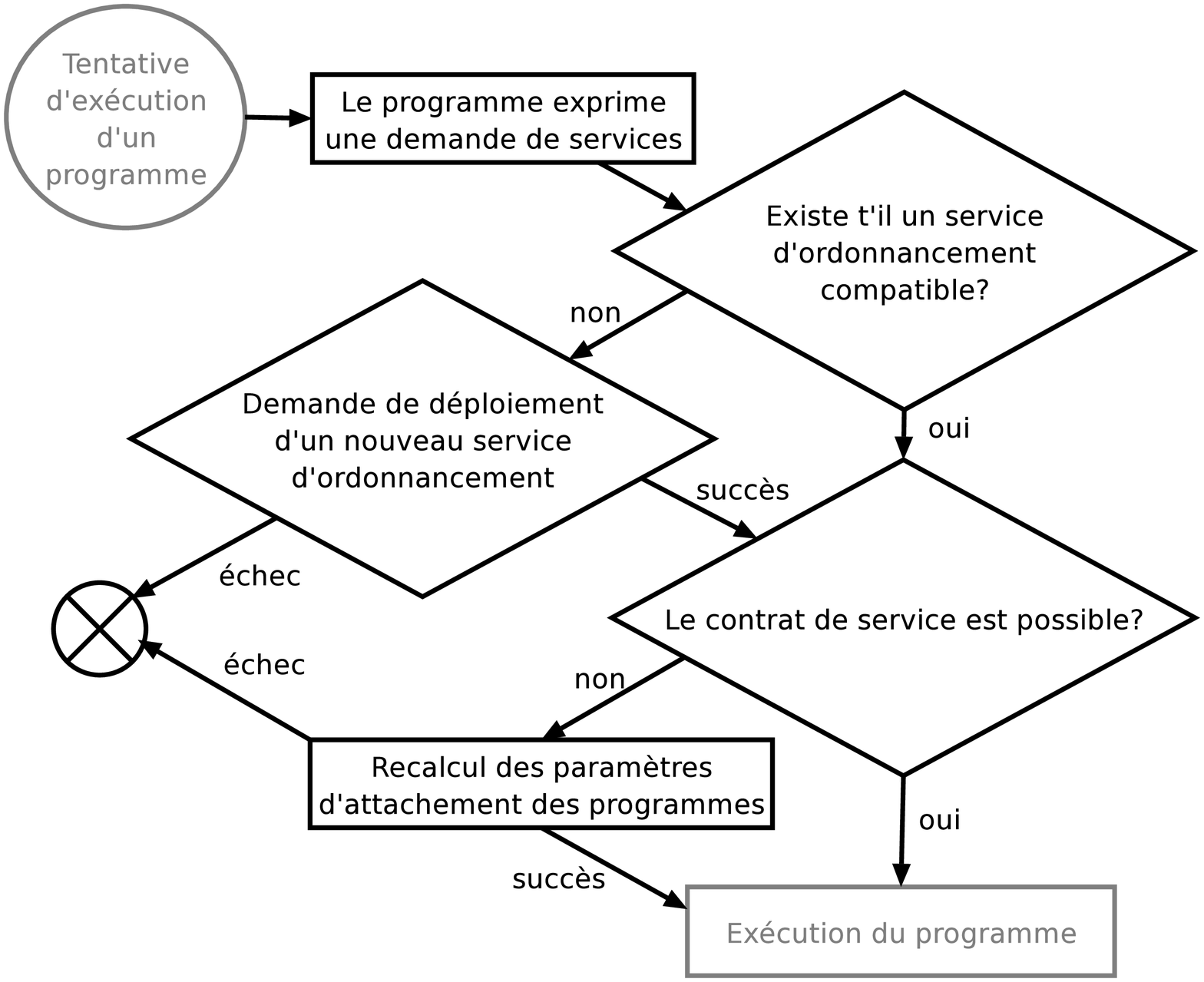}
\caption{\label{schm1} \footnotesize Algorithme de déploiement d'un ordonnanceur}
\end{center}
\end{figure}
\end{center}

%Dans le cadre du déploiement d'une application avec son ordonnanceur, les travaux  de Regehr et Stankovic permettent, d'adresser partiellement i) le problème de négociation de l'intégration d'un ordonnanceur en offrant une classification des ordonnanceurs par type de service d'ordonnancement et ii) le problème de reconfiguration dynamique de la hiérarchie en proposant un algorithme de composition de hiérarchie d'ordonnanceurs.
Le modèle de déploiement d'ordonnanceurs proposé par Regehr et Stankovic
%propose quelques solutions intéressantes. Toutefois, il
 ne permet pas d'établir une demande de ressources au niveau global comme le montre la figure \ref{fig:local}.
 Aussi, la ressource processeur concernée par ce contrat est limitée à la ressource disponible localement au niveau de l'ordonnanceur.
 De plus, ce modèle ne répond pas aux problèmes de négociation de l'intégration d'un ordonnanceur spécifique. %Pour cela, nous proposons un protocole précis d'interaction entre les différents services d'ordonnancement et une application.
  Enfin, l'algorithme de construction interrompt la réalisation de la hiérarchie s'il est impossible de respecter l'un des contrats. Par conséquent, nous proposons:

%ces travaux ne répondent pas aux problèmes de négociation de l'intégration d'un ordonnanceur spécifique en proposant un protocole précis d'intéraction entre les différents services d'ordonnancement et une application.
%De plus, la composition d'une hiérarchie d'ordonnanceurs ne gère pas d'un point de vue global l'harmonisation des besoins en ressource processeur. Par conséquent, nous proposons:
\begin{itemize}
\item de mettre en place un protocole pour autoriser ou non une application à charger son ordonnanceur. Ce protocole est le suivant : une application souhaitant s'exécuter doit exprimer ses besoins d'ordonnancement en utilisant le modèle sémantique défini par Regehr et Stankovic. Cette expression de besoins est alors utilisée par le système pour rechercher dans la hiérarchie s'il existe un service d'ordonnancement compatible. Dans le cas contraire, le système charge le composant d'ordonnancement spécifique à l'application,
\item d'étendre l'algorithme de composition d'une hiérarchie d'ordonnanceurs utilisé par Regehr et Stankovic en remplaçant la condition d'arrêt de l'algorithme par une tentative de réallocation de la ressource processeur entre tous les ordonnanceurs présents. Cette extension permet au système d'essayer de répondre aux besoins exprimés par l'application.
\end{itemize}
Ces deux étapes du déploiement sont illustrées par la Figure \ref{schm1}.
\vspace*{-3,4cm}
%\vspace{-1}
%pour autoriser ou non l'intégration d'un ordonnanceur dans une hiérarchie. Enfin, l'algorithme de %construction

%définir le mode opératoire d'exécution

%lors de son exécution, une application exprime ses besoins d'ordonnancement en utilisant le modèle sémantique défini par Regehr. , le système recherche dans la hiérarchie s'il existe un service d'ordonnancement compatible. Dans le cas contraire, le système charge le composant d'ordonnancement spécifiques à l'application.

% temps le système tente de répondre aux besoins de l'application.
%Pour ce faire, nous étendons l'algorithme de composition de hiérarchie d'ordonnanceur en remplaçant la condition d'arrêt par une tentative de réallocation de la ressource processeur entre tous les ordonnanceurs présents.

%\input{./solution3.tex}
%--------------------------------------------%
%CONCLUSION
\section{Conclusion}
Dans cet article nous avons présenté le principe du déploiement d'une application avec son ordonnanceur dans un système d'exploitation généraliste et décrit les problèmes engendrés. Pour les problèmes non encore adressés par l'utilisation de l'environnement Bossa nous avons proposé i) de valider la pertinence du chargement d'un nouvel ordonnanceur et ii) de vérifier le maintien des contraintes d'ordonnancement lors du chargement d'un nouvel ordonnanceur dans la hiérarchie.

Nous souhaitons maintenant implémenter ces solutions dans l'environnement Bossa. De plus, nous nous proposons d'étudier les possibilités d'adaptabilité de l'application, afin qu'elle puisse revoir ses besoins d'ordonnancement si ceux-ci ne peuvent être satisfaits.

%\section{Conclusion}
%--------------------------------------------%
%BIBLIOGRAPHIE
\bibliography{DECOR04}

\end{document}